\documentclass[aps,twocolumn,prb,showpacs,amsmath,amssymb,superscriptaddress]{revtex4}
\usepackage{graphicx}
\usepackage{dcolumn}
\usepackage{bm}
\usepackage{amsmath}
\usepackage{color}

\newcommand{\bq} {\mathbf{q}}

\newcommand{\nmi} \mathrm   

\begin{document}

\title{Ground-plane screening of Coulomb interactions in two-dimensional systems: How effectively can one two-dimensional system screen
interactions in another?}

\author{L.H. Ho}
\email{laphang@phys.unsw.edu.au}
\affiliation{School of Physics, University of New South Wales,
Sydney NSW 2052, Australia}
\affiliation{CSIRO Materials Science and
Engineering, P.O. Box 218, Lindfield NSW 2070, Australia}

\author{A.P. Micolich}
\email{mico@phys.unsw.edu.au}
\affiliation{School of Physics, University of New South Wales, Sydney NSW 2052, Australia}

\author{A.R. Hamilton}
\affiliation{School of Physics, University of New South Wales, Sydney NSW 2052, Australia}

\author{O.P. Sushkov}
\affiliation{School of Physics, University of New South Wales, Sydney NSW 2052, Australia}

\date{\today}

\begin{abstract}
The use of a nearby metallic ground-plane to limit the range of the
Coulomb interactions between carriers is a useful approach in
studying the physics of two-dimensional (2D) systems.  This approach
has been used to study Wigner crystallization of electrons on the
surface of liquid helium, and most recently, the insulating and
metallic states of semiconductor-based two-dimensional systems. In
this paper, we perform calculations of the screening effect of one
2D system on another and show that a 2D system is at least as
effective as a metal in screening Coulomb interactions.  We also
show that the recent observation of the reduced effect of the
ground-plane when the 2D system is in the metallic regime is due to
intralayer screening.
\end{abstract}

\maketitle

\section{introduction}

In a two-dimensional electron system (2DES), strong Coulomb
interactions between electrons can lead to exotic phenomena such as
the Wigner crystal
state,~\cite{WignerPR34,CrandallPLA71,GrimesPRL79} the fractional
quantum Hall effect,~\cite{TsuiPRL82,LaughlinPRL83} and the
anomalous 2D metallic state.~\cite{AltshulerPhysE01,AbrahamsRMP01,KravchenkoRPP04}
One route to studying the role played by Coulomb interactions is to
limit their length-scale using a metallic ground-plane located close
to the 2DES.~\cite{PeetersPRB84,WidomPRB88}  This approach was first
used in studies of the melting of the Wigner crystal state formed in
electrons on a liquid He surface.~\cite{JiangSS88,MisturaPRB97} More
recently, it has been used to study the role of Coulomb interactions
in the insulating~\cite{HuangCM06} and metallic~\cite{HoPRB08}
regimes of a 2D hole system (2DHS) formed in an AlGaAs/GaAs
heterostructure.

\begin{figure}
\includegraphics[width=8.5cm]{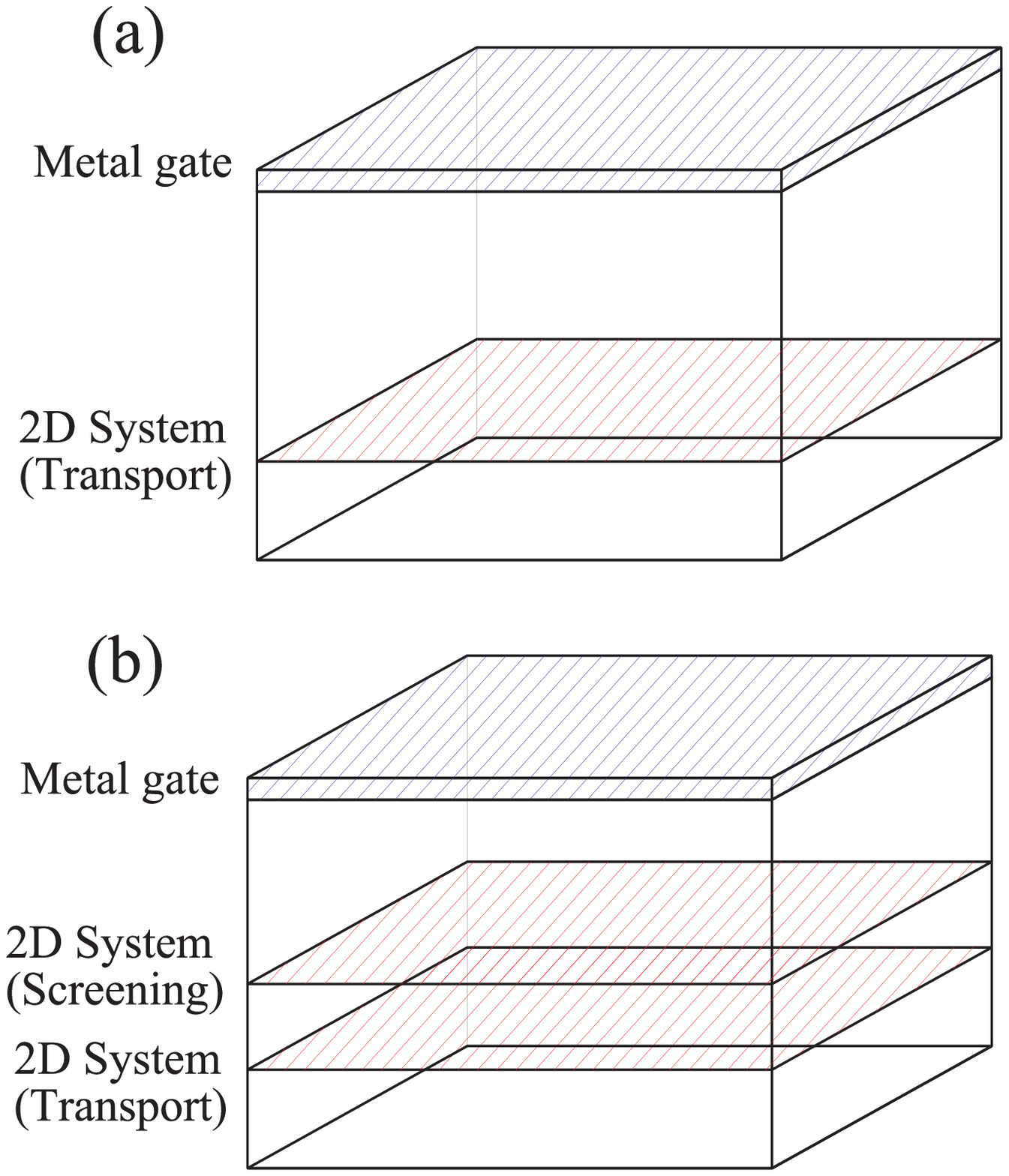}
\caption{\label{fig1}} Schematics of the ground-plane screening
experiments recently performed by (a) Huang {\it et
al.}~\cite{HuangCM06} and (b) Ho {\it et al.}~\cite{HoPRB08}.  In
(b), there are two possible ground-plane configurations.  In the
first, the gate is grounded and the 2D system acts as the
ground-plane.  In the second, the gate is biased to deplete the
upper 2D system, and the gate then acts as the ground-plane instead.
This allows the distance between the transport layer and the
ground-plane to be varied {\it in situ} -- For more details, see
Ref. 14.
\end{figure}

Whereas the study of Coulomb interactions in the insulating
regime~\cite{HuangCM06} was achieved quite straightforwardly using a
metal surface gate separated from the 2DHS by $\sim 500$ nm (see
Fig.~\ref{fig1}(a)), the corresponding study in the metallic regime
could not be achieved in this way.  This is because the higher hole
density $p$ in the metallic regime requires that the distance $d$
between the 2DHS and ground-plane be comparable to the carrier
spacing ($d \sim 2(\pi p)^{-1/2} \sim 50$ nm) to achieve effective
screening, and at the same time that the 2DHS be deep enough in the
heterostructure ($> 100$ nm) to achieve a mobility sufficient to
observe the metallic behavior.  To overcome this challenge, a double
quantum well heterostructure was used (See Fig.~\ref{fig1}(b)) such
that the 2DHS formed in the upper quantum well (screening layer)
served as the ground-plane for the lower quantum well (transport
layer), enabling the measurement of a $\sim 340$ nm deep, high
mobility 2DHS separated by only $50$ nm from a
ground-plane.~\cite{HoPRB08} 

In considering experiments on screening in double quantum well
systems, a natural question to ask is whether a 2D system is as
effective as a metal gate when used as a ground-plane to screen
Coulomb interactions between carriers in a nearby 2D system.  This
is important given that the screening charge in a 2D system is
restricted to two dimensions and the density of states is several
orders of magnitude smaller than in a metal film.  In this paper, we
perform calculations of the screening effect of a ground-plane on a
2D system for two cases: The first where the ground-plane is a metal
and the second where it is a 2D system.  We begin using the
Thomas-Fermi approximation in the absence of intralayer screening in
the transport layer to show that a 2D system is at least as
effective as a metal gate as a ground-plane for the experiment in
Ref.~14.  We also compare the experiments in the
insulating~\cite{HuangCM06} and metallic~\cite{HoPRB08} regimes of a
2D hole system in the Thomas-Fermi approximation, to explain why the
ground-plane has less effect in the metallic regime compared to the
insulating regime.  Finally, since the experiment by Ho {\it et al.}
was performed at $r_{s} > 1$, where the Thomas-Fermi approximation
begins to break down, we extend our model to account for exchange and finite thickness effects to see how these affect the
conclusions from the Thomas-Fermi model.

The paper is structured as follows.  In Section II we derive the
dielectric functions for screening of a 2D hole system by a metal
gate and another nearby 2D hole system.  In Section III, we compare
the various dielectric functions numerically and discuss their
implications for the ground-plane screening experiments of 2D
systems in the insulating~\cite{HuangCM06} and
metallic~\cite{HoPRB08} regimes.  Conclusions will be presented in
Section IV.  For readers unfamiliar with the intricacies of
screening in 2D systems, we give a brief introduction to the
screening theory for a single 2D system in Appendix A to aid them in
understanding the theory developed in Sect.~II. In Appendix B, we
compare our model accounting for exchange and finite
thickness effects to related works on many-body physics in double
quantum well structures.
{
In Appendix C we show how our analysis of ground plane screening can be related to previous work on bilayer 2D systems
that have studied the compressibility. 
In contrast to experiments on ground-plane screening of the interactions within a 2D system,\cite{HuangCM06,HoPRB08} previous experiments on bilayer 2D systems have examined the compressibility of a 2D system by studying penetration of an electric field applied perpendicular to the 2D plane\cite{EisensteinPRB94}. 
We show that although these two screening configurations have quite distinct geometries, our model is consistent with 
widely used analysis\cite{EisensteinPRB94}
describing the penetration of a perpendicular electric field through a 2D system.
}

In the calculations that follow, we use linear screening theory and
the static dielectric function approximation (i.e., $\omega
\rightarrow 0$). Unless otherwise specified, we assume for
convenience that the 2D systems contain holes (electron results can
be obtained with appropriate corrections for charge and mass) to
facilitate direct connection with recent experimental results in
AlGaAs/GaAs heterostructures.~\cite{HuangCM06,HoPRB08}  We also
assume that tunneling between the two quantum wells is negligible
and ignore any Coulomb drag effects (i.e., interlayer exchange and correlations).

\section{The Screening of One 2D Layer by Another}

\begin{figure}
\includegraphics[width=8.5cm]{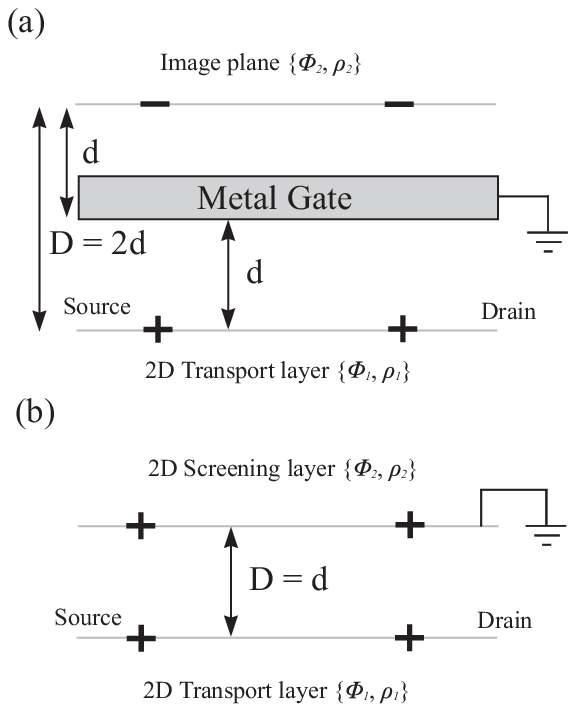}
\caption{\label{fig2}} Schematics showing the two systems considered
in this paper. The transport layer is screened by (a) a metal
surface gate and (b) a second 2D system. In both cases the screening
layer is separated by a distance $d$ from the transport layer, and
the transport (1) and screening (2) layers have independent
potentials $\phi$ and charge densities $\rho$.
\end{figure}

We now begin considering the screening effect of a nearby
ground-plane on a 2D system (transport layer) for two different
configurations.  In the first, the ground-plane (i.e., screening
layer) is a metal surface gate (see Fig.~\ref{fig2}(a)) and in the
second, the ground-plane is another 2D system (see
Fig.~\ref{fig2}(b)). In both cases the transport and screening
layers are separated by a distance $d$.

If we consider some positive external test charge $\rho_{1}^{ext}$
added to the transport layer, this leads to induced charge in both
the transport layer $\rho_{1}^{ind}$ (as in Appendix A) and in the
screening layer $\rho_{2}^{ind}$.  Note however that no external
charge is added to the screening layer, so $\rho_{2}^{ext} = 0$ and
$\rho_{2} = \rho_{2}^{ind}$, whereas $\rho_{1} = \rho_{1}^{ext} +
\rho_{1}^{ind}$.

How we deal with the induced charge in the screening layer differs
in the two cases. In both cases, we consider the transport layer and
a 2nd layer of charge a distance \emph{D} above it; each layer having a potential and
charge density of $\phi_{1}(q)$, $\rho_{1}(q)$ and
$\phi_{2}(q)$, $\rho_{2}(q)$, respectively.
 For a metal surface gate, we can use the standard
image charge approach,~\cite{Jackson99} which involves considering
the induced charge in the screening layer as a 2D layer of negative
`image' charges located a distance $D = 2d$ away from the transport
layer.  This results in $\rho_{2}^{ind} = -\rho_{1}$ for a metal
gate.  The image charge approach assumes that the ground-plane is a
perfect metal.
This assumption is relatively well
satisfied by a typical metal surface gate (Au gate $\sim150$ nm
thick) but not by a 2D system.  Thus when a 2D system is used as the
screening layer, we cannot assume an added induced negative image charge
as we can for the metal.  
Instead, we account for screening by a 2D system by directly calculating
its induced charge, which is located in the screening layer (i.e., at a distance $D
= d$). 
This results in $\rho_{2}^{ind} =
\chi^{0}\phi_{2}$ for a 2D screening layer, where $\chi^{0}$ is the
polarizability, which describes how much $\rho^{ind}$ is produced in
response to the addition of $\rho^{ext}$. Note that we are only considering
the case where the both 2D layers are of the same type of charge (e.g., a bilayer
2DHS).

Charge in one layer leads to a potential in the other via the
interlayer Coulomb interaction:~\cite{Gradshteyn93}

\begin{align}
\label{eqn1} U(q) = \mathcal{F}[\frac{1}{4 \pi \epsilon \sqrt{r^{2}
+ D^{2}}}] = e^{-qD}V(q)
\end{align}

\noindent where $V(q) = \frac{1}
{2\epsilon q}$ is the intralayer Coulomb
interaction, $D$ is the distance between the two layers, and
$\mathcal{F}$ is the Fourier transform.  If the screening layer is a
metal $D = 2d$, and $D = d$ if it is a 2D system. The resulting
potential in the transport layer then becomes:

\begin{eqnarray}
\label{eqn2} \phi_1(\bq) = V(q)\rho_1 + U(q)\rho_2^{ind}(\bq)
\end{eqnarray}

\noindent We discuss how to obtain $\rho_2^{ind}(\bq)$ in section II. A.
The effectiveness of the screening is obtained from the modified 
dielectric function $\epsilon(q,d)$, which we define as the inverse of
the ratio of the screened potential to the unscreened potential for the transport layer:

\begin{align}
\label{eqn4} \epsilon(q,d) = (\frac{\phi_{1}}{\phi_{1}^{ext}})^{-1}
\end{align}

The dielectric function for the transport layer can be obtained for
three possible configurations: no screening layer (i.e., just a
single 2D system), a 2D screening layer and a metal screening layer,
which we denote as $\epsilon_\nmi{single}(q)$, $\epsilon_\nmi{2D}(q,d)$ and
$\epsilon_\nmi{metal}(q,d)$, respectively. 
{
$\epsilon_\nmi{single}(q)$ is the conventional
dielectric function for a single 2D system, and can be recovered by taking the limit $d \rightarrow \infty$ 
for either $\epsilon_\nmi{2D}(q,d)$ or $\epsilon_\nmi{metal}(q,d)$, as we show in Section II Part C. 
If one wanted to take transport theory for a single layer, and consider the effects of ground plane screening, it is only necessary to
replace $\epsilon_\nmi{single}(q)$ with 
$\epsilon_\nmi{2D}(q,d)$ or $\epsilon_\nmi{metal}(q,d)$ to take into account the screening effect of a 2D or metal ground plane, respectively.}

We will now obtain the various $\epsilon$
using an approach involving the Random Phase Approximation (RPA)
\cite{BohmPR53} and Thomas-Fermi (TF)
approximation.~\cite{ThomasFermi27}  At first we will ignore
(Sect.~IIA) and later include (Sect.~IIB) intralayer screening in
the transport layer in the calculations.  Finally, in Sect.~IIC, we
will extend the model for a 2D screening layer to account for its
behavior at lower densities, to confirm that our conclusions from
the simpler calculations are robust.

\subsection{No Intralayer Screening in the Transport Layer}

We begin by considering the case where there is no intralayer
screening in the transport layer.  This is useful because it allows
a straightforward comparison of the effectiveness of the 2DHS as a
ground-plane, without the obscuring effect of intralayer screening.
 To do this calculation, we set $\rho_{1}^{ind} = 0$, such that
$\rho_{1} = \rho_{1}^{ext}$.  In other words, there is only external
charge in the transport layer and only induced charge in the
screening layer.

Considering the metal gate first, we have $\rho_{2}^{ind}(\bq) =
-\rho_{1}(\bq) = -\rho_{1}^{ext}(\bq)$ from the method of images.
If we combine the two results above for $\rho_{1}$ and
$\rho_{2}^{ind}$ with Eqn.~\ref{eqn2}, we obtain:

\begin{align}
\label{eqn5} \phi_{1}(\bq) = (V(q) - U(q))\rho_{1}^{ext}(\bq)
\end{align}

\noindent After using Eqn.~\ref{eqn1} to eliminate $U(q)$,
Eqn.~\ref{eqn4} then gives the dielectric function for the metal
gate:

\begin{align}
\label{eqn6} \frac{1}{\epsilon_\nmi{metal,ns}(q,d)} = 1 - e^{-2qd}
\end{align}

\noindent where the additional subscript $ns$ denotes that
intralayer screening has been ignored.

For a 2D screening layer, the dielectric function is obtained
self-consistently through the RPA as follows.  The induced charge is
related to the screening layer potential by:

\begin{align}
\label{eqn3} \phi_2(\bq) = U(q)\rho_1 + V(q)\rho_2^{ind}(\bq)\\
\label{eqn7} \rho_{2}^{ind}(\bq) = \chi^{0}_{2}(q)\phi_{2}(\bq)
\end{align}

\noindent where $\chi^{0}_{2}$ is the polarizability of the
screening layer, normally given by the 2D Lindhard
function.~\cite{SternPRL67} When this is combined with
Eqn.~\ref{eqn3}, knowing that $\rho_{1} = \rho_{1}^{ext}$, we
obtain:

\begin{align}
\label{eqn8} \rho_{2}^{ind} =
\frac{\chi^{0}_{2}(q)V(q)}{1-\chi^{0}_{2}(q)V(q)}e^{-qd}\rho_{1}^{ext}
\end{align}

\noindent This result is substituted into Eqn.~\ref{eqn2}, and using
Eqns.~\ref{eqn1} and \ref{eqn4} gives:

\begin{align}
\label{eqn9}
\frac{1}{\epsilon_\nmi{2D,ns}(q,d)}=1+\frac{\chi^{0}_{2}(q)V(q)}{1-\chi^{0}_{2}(q)V(q)}e^{-2qd}
\end{align}

\noindent To simplify this expression, we use the Thomas-Fermi
approximation $\chi^{0}_{2}(q) = -e^{2}\frac{dn}{d\mu}$, where
$\frac{dn}{d\mu}$ is the thermodynamic density of states of the 2D
system,~\cite{Davies98} to give:

\begin{align}
\label{eqn10} \frac{1}{\epsilon_\nmi{2D,ns}(q,d)}=1-\frac{q^{TF}}{q +
q^{TF}}e^{-2qd}
\end{align}

\noindent where the Thomas-Fermi wavevector $q^{TF} =
\frac{m^{*}e^{2}}{2\pi\epsilon_{0}\epsilon_{r}\hbar^{2}}$.  Note
that if we take the 2D screening layer to the metallic limit, in
other words, we give it an infinite density of states, which
corresponds to $q^{TF} \rightarrow \infty$, then Eqn.~\ref{eqn10}
reduces to Eqn.~\ref{eqn6}, as one would expect.

\subsection{With Intralayer Screening in the Transport Layer}

We now consider the case where there is intralayer screening (i.e.,
finite polarizability and induced charge) in the transport layer. To
approach this problem, we again place an external charge density
$\rho_{1}^{ext}$ in the transport layer, but now we have induced
charge in both the transport $\rho_{1}^{ind}$ and screening
$\rho_{2}^{ind}$ layers.  Additionally, we label the polarization
$\chi^{0}_{i}(q)$ and Thomas-Fermi wavenumber $q_{i}^{TF}$ where $i
= 1$ or $2$ corresponding to the transport and screening layers,
respectively.  The derivation proceeds as before, but with the
addition of the induced charge density
$\rho_{1}^{ind}(\bq)=\chi^{0}_{1}(q)\phi_{1}(\bq)$ in the transport
layer.  The results obtained are:

\begin{eqnarray}
\label{eqn11} \frac{1}{\epsilon_\nmi{metal,s}(q,d)} &=& \frac{1-e^{-2qd}}{1- V(q)\chi^{0}_{1}(q)(1-e^{-2qd})} \notag\\
\label{eqn12} &=& \frac{1-e^{-2qd}}{1 +
\frac{q_{1}^{TF}}{q}(1-e^{-2qd})}
\end{eqnarray}

\noindent for the metal gate, where the added subscript $s$ denotes
that intralayer screening has been included, and:

\begin{widetext}
\begin{eqnarray}
\label{eqn13} \frac{1}{\epsilon_\nmi{2D,s}(q,d)} &=&
\frac{1-V\chi^{0}_{2}[1 - e^{-2qd}]}{[1-V\chi^{0}_{1}][1 -
V\chi^{0}_{2})]-V^{2}\chi^{0}_{1}\chi^{0}_{2}e^{-2qd}} = \frac{1 +
\frac{q_{2}^{TF}}{q}[1-e^{-2qd}]}{(1+\frac{q_{1}^{TF}}{q})(1+\frac{q_{2}^{TF}}{q})-\frac{q_{1}^{TF}q_{2}^{TF}}{q^{2}}e^{-2qd}}
\end{eqnarray}
\end{widetext}

\noindent when the screening layer is a 2D system.

We can check the consistency of these equations with those in
Sect.~IIA in three ways.  Firstly, by setting $q_{1}^{TF} = 0$,
which corresponds to no screening or induced charge in the transport
layer, Eqns.~\ref{eqn12} and \ref{eqn13} reduce to Eqns.~\ref{eqn6}
and \ref{eqn10}, respectively.  Secondly, if we set $q_{2}^{TF} = 0$
instead, which corresponds to no screening or induced charge in the
screening layer, then $\epsilon_\nmi{2D,s}^{-1}$ in Eqn.~\ref{eqn13}
reduces to $\epsilon_\nmi{single}^{-1}$, which is given in the
Thomas-Fermi approximation by:~\cite{Davies98}

\begin{align}
\label{eqn14} \frac{1}{\epsilon_\nmi{single}(q)} = (1 + \frac{q_{1}^{TF}}{q})^{-1}
\end{align}

\noindent Finally, if we set $q_{2}^{TF} \rightarrow \infty$ to take
the 2D screening layer to the metallic limit, then
$\epsilon_\nmi{2D,s}^{-1}$ in Eqn.~\ref{eqn13} reduces to
$\epsilon_\nmi{metal,s}^{-1}$ in Eqn.~\ref{eqn12}.

\subsection{More Accurate Calculations for 2D Systems at Lower Densities}

Following our relatively simple treatment of ground-plane screening
above, it is now interesting to ask how the results of our
calculations change if we extend our model to account for two
phenomena ignored in our Thomas-Fermi model: exchange 
effects at low densities, and the finite thickness of
the screening and transport layers.

The Thomas-Fermi approximation works well when the interaction
parameter $r_s = (a_{B}^{*} \sqrt{\pi p})^{-1} \lesssim 1$, where
$a_{B}^{*} = 4\pi\epsilon\hbar^{2}/m^{*}e^{2}$ is the effective Bohr
radius.  However, it is not as accurate for 2D systems at lower
densities, such as those used in our experiment \cite{HoPRB08},
where the interaction parameters for the screening and transport
layers were $r_{s} \sim 10$ and $10.2 < r_{s} < 14.3$, respectively.
At such low densities, it is essential to include the effects of
exchange, and a better approximation involves the addition of
the local field correction\cite{Giuliani05}, 
{
which we will use in this section. 
%

One might question if linear screening theory is valid at the large $r_s$ values considered here, or whether it is necessary to 
include non-linear screening effects.
Recent compressibility measurements of 2D systems across the apparent metal-insulator
transition\cite{AllisonPRL06} have shown that the divergence of the inverse compressibility at low densities is 
consistent with non-linear screening 
theories,\cite{ShiPRL02, FoglerPRB2004} which take into account a percolation 
transition in the 2D system. This consistency suggested that at sufficiently low densities, non-linear theories were required to 
adequately describe screening. However, although we are interested in fairly high $r_s$, our analysis relates to measurements performed on the metallic side of the apparent metal-insulator transition, where the effect of inhomogeneities is minimal.  
Indeed, Ref.~25 shows a comparison of the non-linear screening theories\cite{ShiPRL02, FoglerPRB2004} with a standard uniform screening theory.  These theories all yield the same results in our density range of interest, confirming that it is safe to treat the 2D system as homogenous and use linear screening in our calculations.

}

Returning to the inclusion of the local field correction to our calculations;
when considering the case of two 2D layers, we will use the single layer local field
factor $G(\bq)$ to account for intralayer exchange effects, 
and for simplicity, ignore any corresponding interlayer
effects. In this work we will use the Hubbard approximation for $G(\bq)$ 
(see Eqn.~\ref{eqn29} in  Appendix A).
This leads to:

\begin{align}
\label{eqn17} \chi_{i}(q) = \frac{\chi^{0}_{i}(q)}{1 -
V(q)\chi^{0}_{i}(q)[1 - G_{i}(q)]}
\end{align}

\noindent where $G_{i}(q)$ is the local field factor for layer $i =
(1,2)$, respectively. The calculations proceed as before in
Sect.~IIB, except that where we consider intralayer screening in the
transport layer, we have:

\begin{align}
\label{eqn18} \rho^{ind}_{1}(\bq) = \chi_{1}(\bq)
[\phi^{ext}_{1}(\bq) + U(\bq)\rho^{ind}_{2}(\bq)]
\end{align}

\noindent and where the ground plane is a 2D layer, we have:

\begin{align}
\label{eqn19} \rho^{ind}_{2}(\bq) = \chi_{2}(\bq)
[U(\bq)\rho^{ext}_{1}(\bq) + U(\bq)\rho^{ind}_{1}(\bq)]
\end{align}

It is also important to account for the finite thickness of the
screening and transport layers, which are confined to $20$ nm wide
quantum wells in Ref.~14.  To do this, we introduce a
form factor $F(q)$ that modifies the bare Coulomb interaction such
that $V(q) \rightarrow V(q)F(q)$.~\cite{AndoRMP82}  The form factor
is defined as $F(q)= \int \int |\psi(z)|^{2} |\psi(z^{'})|^{2}
e^{-q|z-z^{'}|}dz dz^{'}$, where $\psi(z)$ is the wavefunction of an
electron/hole in the direction perpendicular to the plane of the
quantum well.~\cite{SternJJAP74}  Assuming an infinite-square
potential for the quantum well, we
obtain:~\cite{PricePRB84,GoldPRB87}

\begin{equation}
\label{eqn22} F(q)=\frac{1}{4\pi^{2}+a^{2}q^{2}}[3aq +
\frac{8\pi^{2}}{aq} -
\frac{32\pi^{4}(1-e^{-aq})}{a^{2}q^{2}(4\pi^{2} + a^{2}q^{2})}]
\end{equation}

\noindent where $a$ is the width of the well.  We thus obtain the
dielectric functions $\epsilon(q,d)$ as defined in Eqn.~\ref{eqn4},
where $\phi_1(\bq)$ remains as defined in Eqn.~\ref{eqn2}, giving:

\begin{align}
\label{eqn20}
\frac{1}{\epsilon_\nmi{2D,ns,xf}(q,d)}=1+\frac{\Upsilon_{2}e^{-2qd}}{1-\Upsilon_{2}[1-G_{2}(q)]}
\end{align}

\begin{align}
\label{eqn20b}
\frac{1}{\epsilon_\nmi{metal,s,xf}(q,d)} = \frac{(1-e^{-2qd})[1+\Upsilon_1 G_1(q)]}{1-\Upsilon_1(1-G_1(q) - e^{-2qd})}
\end{align}

\begin{widetext}
\begin{eqnarray}
\label{eqn21} \frac{1}{\epsilon_\nmi{2D,s,xf}(q,d)} = \frac{1 +
\Upsilon_{1}G_{1}(q) - \Upsilon_{2}[1 - G_{2}(q) - e^{-2qd}] +
\Upsilon_{1}\Upsilon_{2}[G_{1}(q)G_{2}(q) - G_{1}(q)(1-e^{-2qd})]}
{[1-\Upsilon_{1}(1-G_{1}(q))][1 -
\Upsilon_{2}(1-G_{2}(q))]-\Upsilon_{1}\Upsilon_{2}e^{-2qd}}
\end{eqnarray}
\end{widetext}

\noindent where $\Upsilon_{i} = \chi^{0}_{i}(q)F(q)V(q)$ and the
additional subscript $xf$ indicates the inclusion of exchange
and finite thickness effects.  As a consistency check,
if we take the 2D screening layer to the metallic limit, by using
$G_{2}(q) = 0$, $\Upsilon_{2} = -q^{TF}_{2}/q$ and the limit
$q^{TF}_{2} \rightarrow  \infty$, and return to zero thickness $F(q) = 1$,
then $\epsilon_{2D,ns,xf}$ in Eqn.~\ref{eqn20} reduces to
$\epsilon_{metal,ns}$ in Eqn.~\ref{eqn6}, and $\epsilon_{2D,s,xf}$
in Eqn.~\ref{eqn21} reduces to $\epsilon_{metal,s}$ in
Eqn.~\ref{eqn12}.  If we separate the two layers by taking $d
\rightarrow \infty$ then both $\epsilon_{2D,s,xf}$ in Eqn.~\ref{eqn21} 
and $\epsilon_{metal,s,xf}$ in Eqn.~\ref{eqn20b} 
reduces to $\epsilon_{single}$ in Eqn.~\ref{eqn14}.  We compare this
work with related studies by Zheng and MacDonald~\cite{ZhengPRB94}
in Appendix B.

\section{Results and Discussion}

In this Section, we will use the various dielectric functions
derived in Sect.~II to answer an important physical question
regarding recent experiments on screening long-range Coulomb
interactions in 2D systems: Does a 2D system screen as effectively
as a metal when used as a ground-plane?

We will answer this question in three stages.  First we will
consider the simplest possible case where there is no intralayer
screening in the transport layer and the Thomas-Fermi approximation
holds.  Our results at this stage are directly applicable to
ground-layer screening studies of dilute 2D systems, such as those
investigating Wigner crystallization on liquid helium
\cite{JiangSS88,MisturaPRB97} and the 2D insulating state in an
AlGaAs/GaAs heterostructure.~\cite{HuangCM06}  They may also be
relevant to recent studies of the metal-insulator transition in Si
MOSFETs,~\cite{TracyCM08} where the gate is likely to produce
significant ground-plane screening in the nearby 2DES located $<40$
nm away, for example.  Second, we will then look at what happens
when intralayer screening is introduced to the transport layer. This
will allow us to understand why the ground-plane has such a
significant effect on the insulating state in the experiment by
Huang {\it et al.} \cite{HuangCM06} and such little effect on the
metallic state in the experiment by Ho {\it et al}.~\cite{HoPRB08}
Finally, since the experiments in Refs.~\cite{HuangCM06} and
\cite{HoPRB08} were performed at $r_{s} >> 1$, we will investigate
how our results change if we extend beyond the Thomas-Fermi
approximation and begin to account for finite thickness and exchange
effects.

\subsection{Thomas-Fermi approximation in the absence of intralayer screening}

To get an understanding of the basic physics of our ground-plane
screening model, we will begin by ignoring any effects of intralayer
screening in the transport layer and use the Thomas-Fermi
approximation to obtain the polarizability $\chi(\bq)$.  There are
two important parameters in our equations: the layer separation $d$
and the wave-number $q$, and to simplify our analysis we will make
these parameters dimensionless by using $q/q^{TF}$ and $d^{TF} = d
\times q^{TF}$ hereafter.  The Thomas-Fermi wave-number $q^{TF}$
contains all of the relevant materials parameters involved in the
experiment. In Ref.~14, where measurements were
performed using holes in GaAs, $\epsilon_{r} = 12.8$ and $m^{*} =
0.38 m_{e}$, giving $q^{TF} = 1.12 \times 10^{9} m^{-1}$ (i.e.,
$(q^{TF})^{-1} = 0.89$ nm).  The corresponding values for electrons
with $m^{*} = 0.067 m_{e}$ are $q^{TF} = 1.97 \times 10^{8} m^{-1}$
and $(q^{TF})^{-1} = 5.06$ nm).  Table 1 presents the $d$ values
corresponding to the four $d^{TF}$ values that we will discuss in
Sects.~IIIA/B.  The first two values correspond to $d = 50$ nm for
holes and electrons, the remaining two allow us to demonstrate what
happens as the screening layer gets much closer to the transport
layer in both cases.

\begin{table}
\caption{\label{table1} The $d$ values for holes and electrons
corresponding to the four $d^{TF}$ values considered in Sections III
A and B.}
\begin{ruledtabular}
\begin{tabular}{ldddd}
$d^{TF}$ & 56.1 & 9.89 & 3 & 1\\
$d_{holes}$ (nm) & 50 & 8.80 & 2.67 & 0.89\\
$d_{electrons}$ (nm) & 283.87 & 50 & 15.18 & 5.06\\
\end{tabular}
\end{ruledtabular}
\end{table}

\begin{figure}
\includegraphics[width=8.5cm]{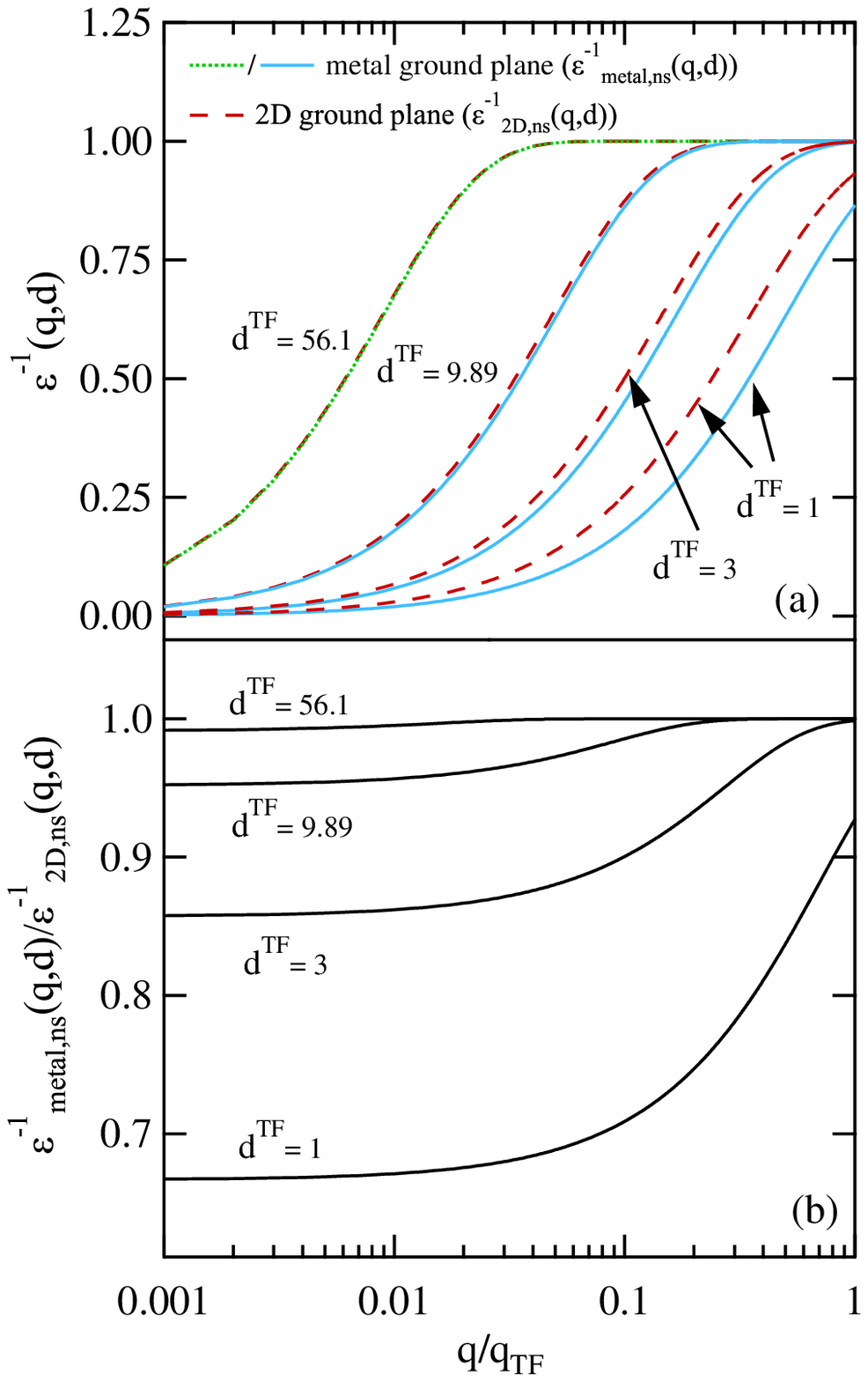}
\caption{\label{fig3}} (a) The inverse dielectric function
$\epsilon^{-1}(q,d)$ vs $q/q_{TF}$ with no intralayer screening in the
transport layer. Data is presented for metal (solid blue and dotted
green lines) and 2D (dashed red lines) screening layers for the four
$d_{TF}$ values presented in Table 1.  The metal gate data for
$d_{TF} = 56.1$ appears as a dotted green line as it serves as
reference data for later figures. (b) The relative effectiveness of
the ground-plane screening due to a 2D screening layer compared to a
metal screening layer, as quantified by the ratio
$\epsilon^{-1}_{metal,ns}/\epsilon^{-1}_{2D,ns}$, vs $q/q_{TF}$ for
the four $d_{TF}$ values.
\end{figure}

To facilitate a comparative analysis of the effectiveness of the
screening, in Fig.~\ref{fig3}(a) we plot the inverse dielectric
function $\epsilon^{-1}$ vs $q/q^{TF}$, with both a metal (solid
blue/dotted green lines) and a 2D system (dashed red lines) as the
screening layer, for the four different $d^{TF}$ values listed in
Table 1. 
Note that for the metal screening layer case ($\epsilon_{metal,ns}(q,d)$ in 
Eqn. \ref{eqn6}), we have explicitly parameterized
$q$ and $d$ into $q/q^{TF}$ and $d^{TF}$, in order to plot the metal and 2D screening 
layer cases on the same axes. 
The metal data for $d^{TF} = 56.1$ is presented as dotted
green line as it serves as reference data for later figures.  Note
that $\epsilon^{-1} = 1$ corresponds to no screening, and
$\epsilon^{-1} = 0$ corresponds to complete screening of a test
charge placed in the transport layer. Considering the large $d^{TF}$
limit first, $\epsilon^{-1}$ only deviates from 1 at small
$q/q^{TF}$, and heads towards $\epsilon^{-1} = 0$ as $q/q^{TF}
\rightarrow 0$.  In other words, screening is only effective at
large distances from a test charge added to the transport layer.
This makes physical sense if one considers the electrostatics of
ground-plane screening.  The ground-plane acts by intercepting the
field lines of the test charge such that they are no longer felt in
other parts of the transport layer.  This is only effective at
distances from the test charge that are much greater than the
ground-plane separation $d$, and thus the ground-plane acts to limit
the range of the Coulomb interaction in the transport layer, as
pointed out by Peeters.~\cite{PeetersPRB84}  With this in mind, it
is thus clear why the point of deviation from $\epsilon^{-1} = 1$
shifts to higher values of $q/q^{TF}$ as $d^{TF}$ is reduced.
Indeed, all four lines pass through a common $\epsilon^{-1}$ value
when $q/q^{TF} = \frac{1}{d^{TF}}$ reflecting this electrostatic
aspect of ground-plane screening.


Turning to the central question of the effectiveness of a 2D layer
as a ground-plane, in Fig.~\ref{fig3}(a) it is clear from the
increasing discrepancy between the solid and dashed lines that the
2D system becomes less effective than a metal as $d^{TF}$ is
reduced. To quantify this, in Fig.~\ref{fig3}(b) we plot the ratio
of the two dielectric constants
$\epsilon_\nmi{metal,ns}^{-1}/\epsilon_\nmi{2D,ns}^{-1}$, with a ratio of
$1$ indicating equivalent screening and $< 1$ indicating that a 2D
system is less effective than a metal.  For large separations, for
example $d^{TF} = 56.1$, which corresponds directly to the
experiment by Ho {\it et al.}, a 2D system screens as effectively as
the metal gate to within $1\%$.  However if the screening layer is
brought very close to the transport layer $d^{TF} \sim 1$ (i.e., the
screening layer is only a Thomas-Fermi screening length away from
the transport layer) then the effectiveness of the 2D system as a
ground-plane is reduced to $\sim 66\%$ of that of a metal layer at
an equivalent distance. It is important to note that correlations
between the two layers can be significant for such small
separations, and hence this increasing discrepancy should be
considered as a qualitative result only.  Furthermore, as we will
see in Sect.~IIIC, exchange actually acts to enhance the
effectiveness of the 2D system as a ground-plane, making the
Thomas-Fermi result above a significant underestimate of the true
ground-plane screening of a 2D system in the low density limit.

\subsection{Thomas-Fermi Approximation with Intralayer Screening in the Transport Layer}

\begin{figure}
\includegraphics[width=8.5cm]{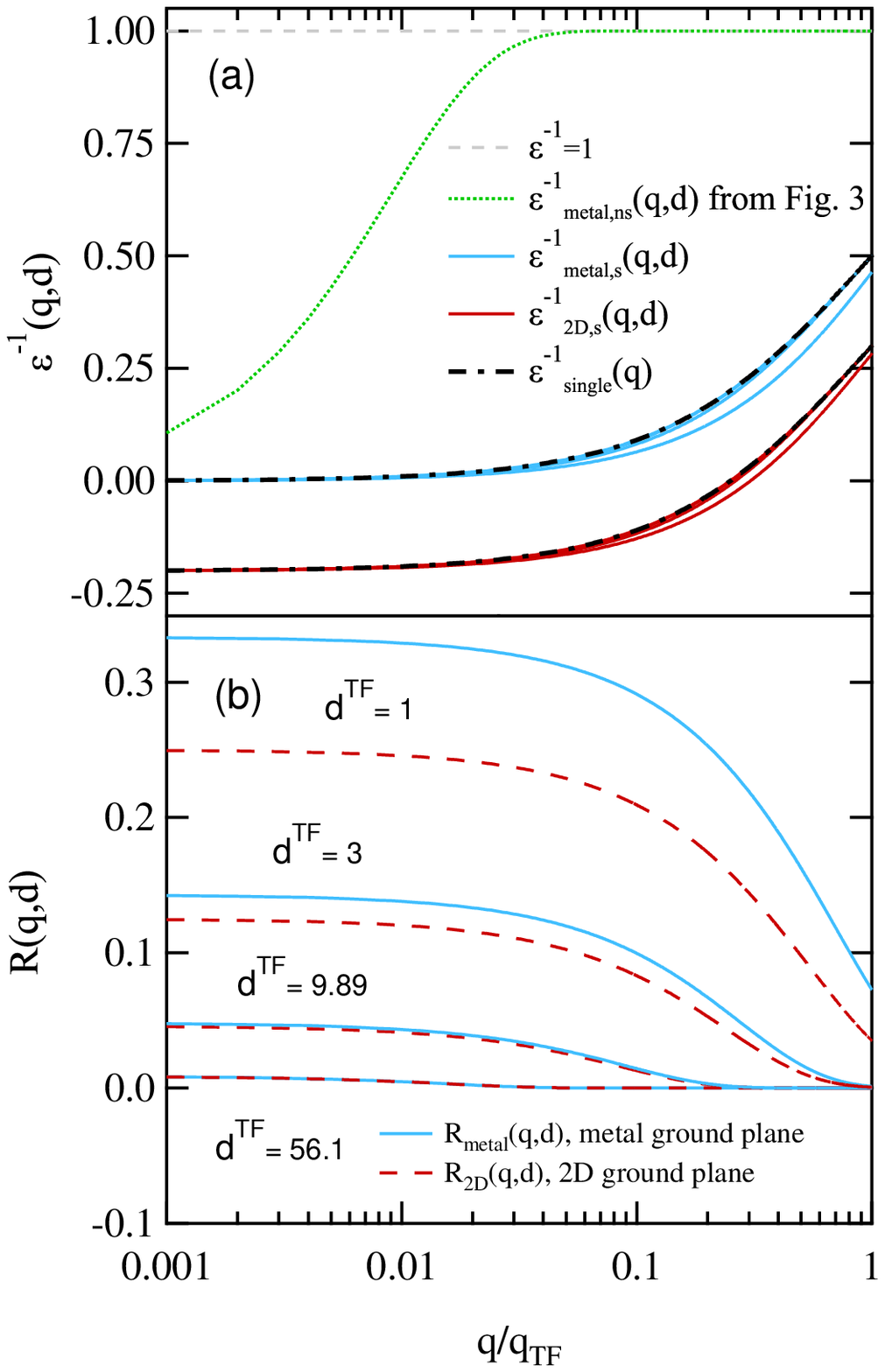}
\caption{\label{fig4}} 
Effect of a screening layer on a 2D system with intralayer screening.
(a) 
{
Firstly, in order to show the relative effects of intralayer and ground plane screening,
we plot the dielectric functions $\epsilon^{-1}=1$ corresponding to no 
intralayer or ground plane screening (grey dashed horizontal line), $\epsilon_\nmi{metal,ns}^{-1}$ with metal screening layer at $d_{TF} = 56.1$ and no intralayer screening (dotted green line, data from Fig.~\ref{fig3}(a)), 
and $\epsilon_\nmi{single}^{-1}$ with intralayer screening but no ground-plane (dash-dotted black line).
We then consider the effect of the metal screening layer when the intralayer screening is included, by plotting 
$\epsilon^{-1}_\nmi{metal,s}(q,d)$ (solid blue lines) for the four values of $d^{TF}$ shown in Table I. Moving through the traces from upper left to lower right corresponds to decreasing $d^{TF}$.
A similar set of curves are shown for the case of a 2D screening layer ($\epsilon^{-1}_\nmi{2D,s}(q,d)$, solid red lines) 
which has been offset vertically by -0.2 for clarity, along with a duplicate of $\epsilon_\nmi{single}^{-1}$ (dash-dotted black line).
}
Since the dielectric functions almost lie on top of each other when intralayer screening is present, in 
(b) we plot $R$, the relative enhancement of $\epsilon^{-1}$ due to the ground plane.
Calculations for the four different values of $d_{TF}$ in Table 1 are shown 
for metal (solid blue lines) and 2D (dashed red lines) ground-planes.
\end{figure}

We now add intralayer screening in the transport layer to our
Thomas-Fermi model, and begin by asking: What is the magnitude of
this intralayer screening contribution, independent of any
ground-plane screening effects? In Fig.~\ref{fig4}(a), we plot the
inverse dielectric function $\epsilon_\nmi{single}^{-1}$ (dash-dotted
black line) for a 2D system with intralayer screening and {\it no}
nearby ground-plane. For comparison, we also show the data from
Fig.~\ref{fig3}(a) for a metal ground-plane with $d_{TF} = 56.1$
(green dotted line) and the expectation with no screening,
$\epsilon^{-1} = 1$ for all $q/q_{TF}$ (grey dashed horizontal line) in
Fig.~\ref{fig4}(a).  It is clear that the addition of intralayer
screening has a very significant impact on the dielectric function,
more so than the addition of a ground-plane.  Indeed, returning to
an electrostatic picture and ignoring exchange and correlation
effects, $\epsilon_\nmi{single}^{-1}$ should assume the $d^{TF}
\rightarrow 0$ limit of $\epsilon_\nmi{2D,ns}^{-1}$, the Thomas-Fermi model in the absence of
intralayer screening.

We now reintroduce the ground-plane, and in Fig.~\ref{fig4}~(a), we
plot the combined screening contributions for metal ($\epsilon^{-1}_\nmi{metal,s}(q,d)$, solid blue
lines) and 2D ($\epsilon^{-1}_\nmi{2D,s}(q,d)$, solid red lines) screening layers. 
{
These are shown for the four different values of $d^{TF}$ in Table 1.}
The values for the
2D system are offset vertically by $-0.2$ for clarity.  The
intralayer screening and ground-plane screening both contribute to
the total screening, albeit on different length scales. This can be
seen by comparing the data in Fig.~\ref{fig4}(a) to that in
Fig.~\ref{fig3}(a), with the intralayer screening clearly the
dominant contribution. 
As a result, distinguishing between individual traces in the sets corresponding to the metal (blue lines) or 2D (red lines) ground planes in Fig.~\ref{fig4}(a) is difficult.
Hence, to better quantify the enhancement that the
ground-plane gives over intralayer screening alone, in
Fig.~\ref{fig4}(b) we plot the relative ground-plane enhancement
$R_\nmi{metal,s} = (\epsilon_\nmi{single}^{-1} -
\epsilon_\nmi{metal,s}^{-1})/|\epsilon_\nmi{single}^{-1}|$ (solid blue
lines) and $R_\nmi{2D,s} = (\epsilon_\nmi{single}^{-1} -
\epsilon_\nmi{2D,s}^{-1})/|\epsilon_\nmi{single}^{-1}|$ (dashed red lines),
respectively.  Note that the ground-plane only provides significant
enhancement over intralayer screening alone as $d^{TF}$ becomes
small, and as in Sect.~IIA, only provides enhancements at small
$q/q^{TF}$.  The small discrepancies between the data for the metal
and 2D screening layers in Figs.~\ref{fig4}(b) directly reflect the
increased effectiveness of the metal ground-plane over a 2D
ground-plane shown in Fig.~\ref{fig3}(b).

The data in Figs.~\ref{fig3} and ~\ref{fig4} provide an interesting
insight into recent experiments on ground-plane screening in 2D hole
systems in the insulating and metallic
regimes.\cite{HuangCM06,HoPRB08}  Due to the low hole density and
conductivity in the insulating regime, intralayer screening is less
effective and the dominant contribution to screening is the
ground-plane, which acts to limit the length scale of the Coulomb
interactions, as Fig.~\ref{fig3}(a) shows.  This results in the
ground-plane having a marked effect on the transport properties of
the 2D system, as shown by Huang {\it et al}.~\cite{HuangCM06}  In
comparison, for the metallic state, where the density and
conductivity are much higher, intralayer screening is the dominant
contribution, and a ground-plane only acts as a long-range
perturbation to the screening, as shown in Fig.~\ref{fig4}(a). This
perturbation to the intralayer screening is particularly small at
$d^{TF} = 56.1$ and results in the ground-plane having relatively
little effect on the transport properties in the metallic regime, as
found by Ho {\it et al}.~\cite{HoPRB08} Although Fig.~\ref{fig4} (b) suggests
that decreasing $d^{TF}$ will increase the effect of the ground plane, in practice 
there are issues in achieving this. For holes in GaAs, there is little scope 
for further reducing $d$ due to increasing Coulomb drag and interlayer tunnelling effects. Also, 
in our model we have neglected interlayer exchange and correlation effects, and 
these may become significant at these lower distances. 


\subsection{Beyond the Thomas-Fermi Approximation}

Following our relatively simple treatment of ground-plane screening
above, it is now interesting to ask how the results of our
calculations change if we extend our model to account for two
phenomena ignored in our Thomas-Fermi model: exchange effects at 
low densities, and the finite thickness of
the screening and transport layers.

The inclusion of the Hubbard local field correction $G(q)$, finite
thickness form-factor $F(q)$, and the use of the Lindhard function for 
$\chi^0(q)$ adds two new parameters to the
analysis, the well thickness $a = 20$ nm and the Fermi wave-vector
$k_{F}$. This removes our ability to reduce the problem down to a
single adjustable parameter $d^{TF}$ as we did in Sects.~IIIA and B.
Additionally, accounting for finite well width puts a lower limit on
$d$, which must be greater than $a$ to ensure that the wells remain
separate.  Hence for the remaining analysis we will only consider $d
= 50$ and $30$ nm, which correspond to $d^{TF} = 56.1$ and $33.7$,
respectively. 
As in earlier
sections, we will first analyse the dielectric function ignoring
intralayer screening in the transport layer, which we achieve by
setting $\rho^{ind}_{1} = 0$.

\begin{figure}
\includegraphics[width=8.5cm]{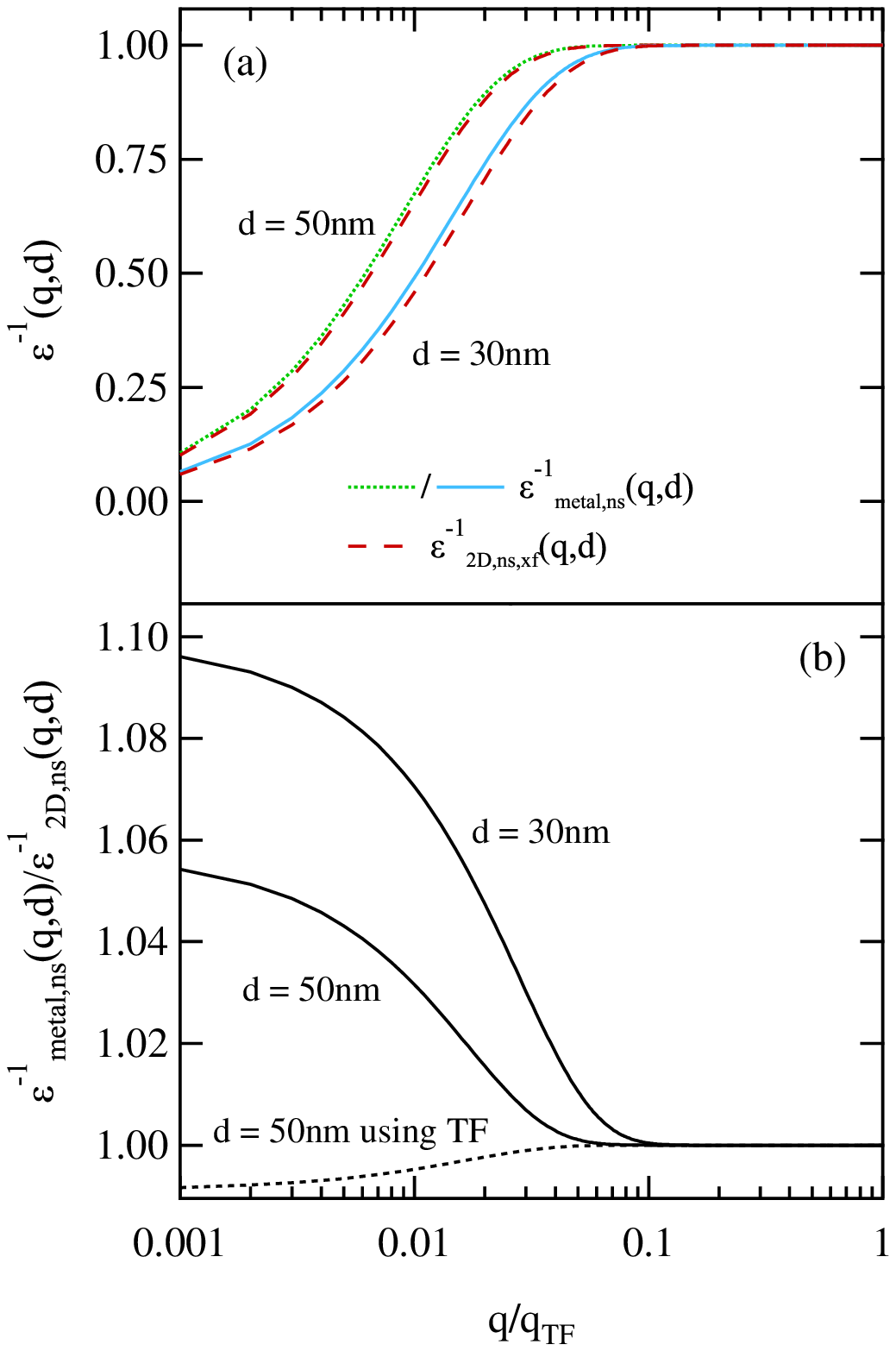}
\caption{\label{fig5}} (a) Plots of $\epsilon^{-1}$ vs $q/q_{TF}$
for a metal (solid blue and dotted green line) and 2D (dashed red
lines) ground plane for $d = 30$ and $50$ nm accounting for exchange
and finite thickness effects but ignoring intralayer screening in
the transport layer. The dotted green line corresponds to that in
Fig.~\ref{fig3}(a).  (b) A plot of the relative screening effect of
a 2D layer compared to a metal (solid lines), as quantified by the
ratio $\epsilon_\nmi{metal}^{-1}/\epsilon_\nmi{2D}^{-1}$. In contrast to the
results for the Thomas-Fermi model (dashed line - data from
Fig.~\ref{fig3}(b)), we find that a 2D layer is actually more
effective than a metal as a ground-plane when exchange
and finite thickness effects are included in the calculation.
\end{figure}

In Fig.~\ref{fig5}(a), we plot $\epsilon_\nmi{2D,ns,xf}^{-1}$ (dashed
red lines) obtained using Eqn.~\ref{eqn21} for $d = 50$ and $30$ nm,
and for comparison, $\epsilon_\nmi{metal,ns}^{-1}$ for $d = 50$ nm
(dotted green line) from Fig.~\ref{fig3}(a) and the corresponding
result for $d = 30$ nm (solid blue line).  One of the more
significant effects of exchange in 2D systems is that it leads to
negative compressibility~\cite{TanatarPRB89,EisensteinPRB94} for
$r_{s} \gtrsim 2$. 
{
It is well known from the field penetration experiments of 
Eisenstein \emph{et al.}\cite{EisensteinPRB94} that negative compressibility is related to an overscreening of the applied electric field, leading to a negative penetration field.
In our calculations, a similar overscreening is observed, with a 2D system producing more effective
ground-plane screening than a metal gate at}
intermediate $q/q^{TF}$ in Fig.~\ref{fig5}(a).  The enhanced
screening when the ground-plane is a 2D system is evident in
Fig.~\ref{fig5}(b), where we plot the ratio
$\epsilon_\nmi{metal,ns}^{-1}/\epsilon_\nmi{2D,ns}^{-1}$, which takes values
greater than $1$ for $q/q^{TF} \lesssim 0.1$.

\begin{figure}
\includegraphics[width=8.5cm]{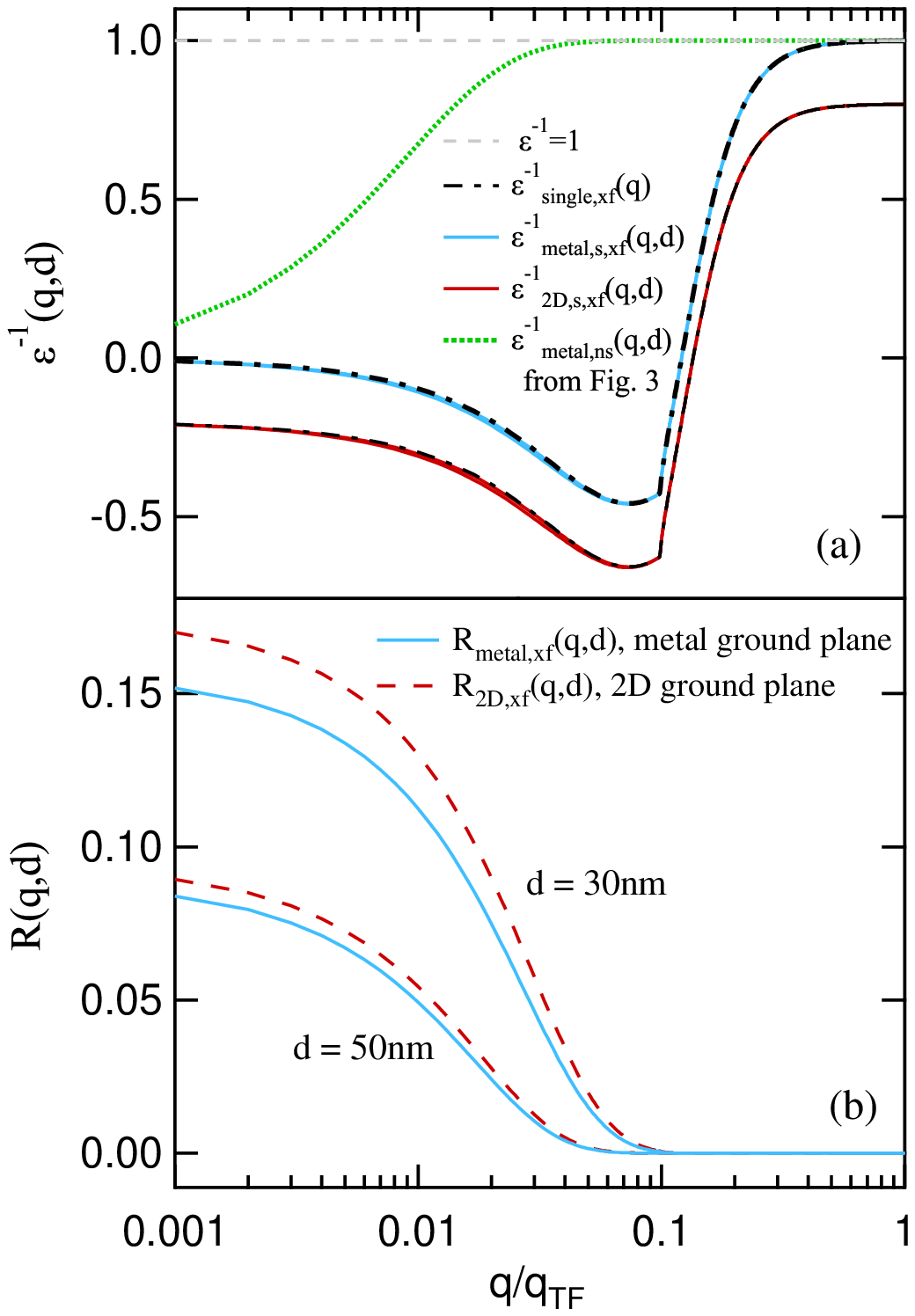}
\caption{\label{fig6}} 
Effect of a screening layer on a 2D system with intralayer screening, with exchange and finite thickness effects included. (a)
{
Firstly, in order to show the relative effects of intralayer and ground plane screening,
we plot the dielectric functions $\epsilon^{-1}=1$ corresponding to no 
intralayer or ground plane screening (grey dashed horizontal line), $\epsilon_\nmi{metal,ns}^{-1}$ with metal screening layer at $d = 50$nm and no intralayer screening (dotted green line, data from Fig.~\ref{fig3}(a)), 
and $\epsilon_\nmi{single,xf}^{-1}$ with intralayer screening but no ground-plane (dash-dotted black line).
We then consider the effect of the metal screening layer when the intralayer screening is included, by plotting 
$\epsilon^{-1}_\nmi{metal,s,xf}(q,d)$ (solid blue lines) for $d=30$ and $50$nm. 
A similar set of curves are shown for the case of a 2D screening layer ($\epsilon^{-1}_\nmi{2D,s,xf}(q,d)$, solid red lines) 
which has been offset vertically by -0.2 for clarity, along with a duplicate of $\epsilon_\nmi{single,xf}^{-1}$ (dash-dotted black line).
Since the dielectric functions almost lie on top of each other when intralayer screening is present, in 
(b) we plot $R$, the relative enhancement of $\epsilon^{-1}$ due to the ground plane.
Calculations for $d=30$ and $50$nm are shown 
for metal (solid blue lines) and 2D (dashed red lines) ground-planes.}

\end{figure}


We now reintroduce intralayer screening in the transport layer, and
in Fig.~\ref{fig6}(a) we plot $\epsilon^{-1}_\nmi{metal,s,xf}$ (solid
blue line) and $\epsilon^{-1}_\nmi{2D,s,xf}$ (dashed red line) for $d =
50$ and $30$ nm. The values for the 2D system are offset vertically by $-0.2$
for clarity.  For comparison, we also plot $\epsilon_\nmi{single}^{-1}$
(dash-dotted black lines --- duplicated and offset vertically by
$-0.2$), along with the data from Fig.~\ref{fig3}(a) for a metal
ground-plane at $d = 50$ nm with no intralayer screening (green
dashed line), and the expectation with no screening $\epsilon^{-1} =
1$ for all $q/q_{TF}$ (grey dashed horizontal line). As we found earlier
with the Thomas-Fermi model (see Fig.~\ref{fig4}(a)), the inclusion
of intralayer screening has a profound effect on the dielectric
function, contributing significantly more to the overall screening
than the addition of a ground-plane does alone.  This demonstrates
the robustness of one of the key results of Sect.~IIIB, namely that
in the metallic regime,~\cite{HoPRB08} where intralayer screening
effects are significant, the ground-plane screening contribution is
overwhelmed by the intralayer screening contribution.  This leads to
a significantly reduced ground-plane effect than one would expect
from studies in the insulating regime.~\cite{HuangCM06}

The effect of including exchange and finite thickness
effects in the calculation is evident by comparing
$\epsilon_\nmi{2D,s,xf}^{-1}$ in Fig.~\ref{fig6}(a) with
$\epsilon_\nmi{2D,s}^{-1}$ in Fig.~\ref{fig4}(a). Considering the
individual contributions, because $F(q) \leq 1$, the finite
thickness of the quantum well acts to reduce the effectiveness of
the 2D layer as a ground-plane.  In contrast, the negative
compressibility produced by the exchange contribution acts to
significantly enhance the screening, and as Fig.~\ref{fig6}(a)
shows, has its most significant impact at intermediate $q/q_{TF}$,
where the dielectric function becomes negative, as discussed by
Dolgov, Kirzhnits and Maksimov \cite{DolgovRMP81}, Ichimaru
\cite{IchimaruRMP82}, and Iwamoto \cite{IwamotoPRB91}. The combined
effect of $G(q)$ and $F(q)$ is to significantly enhance the
screening at intermediate $q/q_{TF}$ whilst reducing it to levels
comparable to the metal ground-plane for large $q/q_{TF}$.  In other
words, the added density-dependence in our Hubbard model leads to
enhanced mid-range screening at the expense of short-range
screening.  A physical interpretation for this behavior is that at
low densities there are insufficient carriers available to screen
effectively close to a test charge, whilst at intermediate ranges,
the negative compressibility produced by exchange leads to a higher
availability of carriers and better screening than there would
otherwise be at higher carrier densities where exchange is not as
significant. It is also interesting to consider why the introduction
of exchange and finite thickness effects have such a
profound effect on the intralayer screening contribution compared to
the ground-plane screening contribution.  This occurs because the
impact of $G(q)$ and $F(q)$ on the ground-plane contribution is
strongly attenuated by the $e^{-2qd}$ terms that appear in
Eqns.~\ref{eqn20} and \ref{eqn21}. Such terms don't occur for the
intralayer screening contribution, which significantly enhances the
impact of the negative compressibility, as is clear by comparing
Fig.~\ref{fig6}(a) with Fig.~\ref{fig5}(a).

We close by considering the relative effectiveness of the metal
and 2D ground-planes with all considerations included in the
calculations.  In Fig.~\ref{fig6}(b) we plot the relative
ground-plane enhancements $R_\nmi{metal,s,xf} = (\epsilon_\nmi{single,xf}^{-1}
- \epsilon_\nmi{metal,s,xf}^{-1})/|\epsilon_\nmi{single,xf}^{-1}|$ (solid blue
lines) and $R_\nmi{2D,s,xf} = (\epsilon_\nmi{single,xf}^{-1} -
\epsilon_\nmi{2D,s,xf}^{-1})/|\epsilon_\nmi{single,xf}^{-1}|$ (dashed red
lines) for $d = 50$ and $30$ nm. As in Fig.~\ref{fig5}(b), we find
that exchange, finite thickness and intralayer
screening result in the 2D ground-plane screening {\it more}
effectively than a metal ground-plane, with the difference between
the two becoming greater as $d$ is decreased. For $d = 50$ nm, the
ground-plane separation used in Ref.~14, the
ground-plane has significantly more effect ($\sim 8-9\%$) than it
does in the more simple Thomas-Fermi model ($\sim 1\%$) presented
earlier.

\section{Summary of Results and Comparison with experiment}

We have performed theoretical calculations to investigate the
relative effectiveness of using a metal layer and a 2D system as a
ground-plane to screen Coulomb interactions in an adjacent 2D
system.  This is done for two cases: the first is the relatively
simple Thomas-Fermi approximation, and the second is the Hubbard
approximation, where we account for exchange and also finite
thickness effects.  This study was motivated by recent experiments
of the effect of ground-plane screening on transport in
semiconductor-based 2D systems.

There were three key findings to our study.  Firstly, a 2D system is
effective as a ground-plane for screening Coulomb interactions in a
nearby 2D system, which was an open question following the recent
experiment by Ho {\it et al}.~\cite{HoPRB08}  In the Thomas-Fermi
approximation, a metal and a 2D system are almost equally effective
at screening the long-range Coulomb interactions in the nearby 2D
system, with the metal becoming relatively more effective as the
ground-plane separation $d$ is decreased.  

Secondly, our
calculations provide an explanation for why ground-plane screening
has much more effect in the insulating regime than it did in the
metallic regime.  
Due to the low hole density and
conductivity in the insulating regime, intralayer screening is weak
and the dominant contribution to screening is the
ground-plane, which acts to limit the length scale of the Coulomb
interactions. This results in the
ground-plane having a marked effect on the transport properties of
the 2D system, as shown by Huang {\it et al}.~\cite{HuangCM06}
In the metallic regime, intralayer screening
cannot be ignored.  In addition to being the dominant contribution
for long-range interactions (i.e., at small $q$), the intralayer
screening contribution is non-zero over a much wider range of $q$,
turning the ground-plane contribution into little more than a small
change to the overall screening in the 2D system, which is consistent with 
the experiment by Ho {\it et al.} \cite{HoPRB08}. 

Finally, since
both experiments were performed at $r_{s} >> 1$, where the
Thomas-Fermi approximation is invalid, we reconsider our
calculations involving 2D systems using the Hubbard approximation
for the local field correction.  We show that our argument regarding
the physics of ground-plane screening in the metallic and insulating
regimes remains robust, but that exchange effects lead to a 2D
system being {\it more} effective than a metal layer as a
ground-plane.  This is due to the exchange-driven negative
compressibility that occurs\cite{TanatarPRB89} at $r_{s} \gtrsim 2$.

\section{Further Work}

While our results suggest that ground-plane effects on a metallic transport layer
should strengthen as the ground-plane separation $d^{TF}$ is reduced, there
are a number of issues that complicate this argument.  Firstly,
for holes in GaAs, such as the experiment in
Ref.~14, there is little scope to further reduce $d$ due
to the increasing Coulomb drag  and interlayer tunnelling effects
that would result. 
However, it may be
possible to experimentally modify $d^{TF}$ by moving to a different
material system where $(q^{TF})^{-1}$ is larger.  For example, in
InAs \cite{AdachiJAP82}, where $m^{*} = 0.026 m_{e}$ and $\epsilon =
14.6 \epsilon_{0}$, we would have $(q^{TF})^{-1} = 14.9$ nm, or InSb
\cite{GoldammerJCG99} where $m^{*} = 0.0145 m_{e}$ and $\epsilon =
17.7 \epsilon_{0}$ gives $(q^{TF})^{-1} = 32.3$ nm.  These
$(q^{TF})^{-1}$ values are 17 and 28 times larger than those in
Ref.~14, respectively. This would allow us to reduce $d_{TF}$ without changing
$d$, thus avoiding the problems above.

{We note that} our model neglects interlayer exchange and correlation
effects, which may become significant at these {small} distances $d$, as suggested 
by calculations at $r_s=4$ by Liu \emph{et al.} ~\cite{LiuPRB96} It could be interesting to 
investigate the effect of including the interlayer exchange and correlation effects on the ground 
plane screening at small $d$. This may require using better approximations for the local field correction 
such as that developed by  Singwi, Tosi, Land and
Sj\"{o}lander~\cite{SingwiPR68} (STLS), as there is no equivalent to the Hubbard approximation for interlayer local field corrections.


We also note that using the technique in Ref.~14 and the
theory presented here, it would be possible to study the breakdown
of intralayer screening in the transport layer as it is evolved from
the metallic to insulating regime.  This could be compared with
compressibility measurements of a 2D system across the apparent
metal-insulator transition \cite{AllisonPRL06}, possibly providing
new insight into the mechanism driving this transition. 

Lastly, in this paper we only calculate the screening of the ground-plane on
the transport layer via the dielectric function. It would be
interesting to take this work further to calculate the effect of the
ground-plane on the actual carrier transport through the transport
layer. Combining the theory presented here, and various models  
of the metallic and insulating behaviours (see review papers~\cite{AltshulerPhysE01,AbrahamsRMP01,KravchenkoRPP04}), 
it may be possible to determine how each
of the models are affected by the presence of a ground plane, and would allow us 
compare this with the experimental data in more detail.

\section{Acknowledgements}

This work was funded by Australian Research Council (ARC).  L.H.H.
acknowledges financial support from the UNSW and the CSIRO.  We thank M.
Polini, I.S. Terekhov and F. Green for helpful discussions.

\appendix
\section{Brief Review of Screening Theory for a Single 2D System}

In this Section we briefly review the basics of screening in a
single 2D system.  Readers familiar with screening theory may wish
to proceed directly to Sect.~II.  A more extended discussion can be
found in Refs.~\cite{AshcroftSSP76,Davies98,Giuliani05}.

Screening occurs when the carriers in a 2D system reorganize
themselves in response to some added `external' positive charge
density, leading to an electrostatic potential determined by
Poisson's equation.  This reorganization produces a negative
`induced' charge density that acts to reduce or `screen' the
electric field of the external charge.  In proceeding, it is
mathematically convenient to instead treat the problem in terms of
wave-vectors ($q$-space) so that the (intralayer) Coulomb potential
$V(r) = \frac{1}{4\pi\epsilon r}$ becomes $V(q) = \frac{1}
{2\epsilon q}$.~\cite{Davies98}

There are two key parameters of interest in an analysis of
screening. The first is the polarizability $\chi(q)$, which relates
the induced (screening) charge density $\rho^{ind}(\bq)$ to the
external (unscreened) potential $\phi^{ext}(\bq)$:

\begin{align}
\label{eqn25} \rho^{ind}(\bq) = \chi(q)\phi^{ext}(\bq)
\end{align}

\noindent The second is the dielectric function $\epsilon(q)$, which
relates the total (screened) potential $\phi(\bq)$ to the external
(unscreened) potential $\phi^{ext}(\bq)$:

\begin{align}
\label{eqn26} \phi(\bq)=\phi^{ext}(\bq)/\epsilon(q)
\end{align}

\noindent Conceptually, the polarizability describes how much
induced charge density is produced in response to the addition of
the external charge density, hence it is also often called the
density-density response function.~\cite{JonsonJPC76}  The
dielectric function is a measure of how effective the screening is:
$\epsilon^{-1} = 1$ corresponds to no screening and $\epsilon^{-1} =
0$ corresponds to perfect screening.~\cite{DolgovRMP81}  The two
parameters can be linked via $\phi^{ext}$ and the Coulomb potential
$V(q)$, such that:

\begin{align}
\label{eqn27} \frac{1}{\epsilon(q)} = 1 + V(q)\chi(q)
\end{align}

\noindent The results above are precise aside from the assumption of
linear response.  However, continuing further requires calculation
of $\chi(q)$.  This cannot be achieved exactly, and requires the use
of approximations.  In the simplest instances, a combination of the
Thomas-Fermi~\cite{ThomasFermi27,Davies98} (TF) and Random Phase
Approximations~\cite{BohmPR53} (RPA) can be used.  However, to
properly account for exchange and/or correlation, particularly at lower
carrier densities, more sophisticated approximations, such as those
developed by Hubbard~\cite{HubbardPRSLA58} or Singwi, Tosi, Land and
Sj\"{o}lander~\cite{SingwiPR68} (STLS) should be used.  For a single
2D layer, this leads to a correction to the induced charge:

\begin{align}
\label{eqn28} \rho^{ind}(\bq) = \chi^0(q)[\phi^{ext}(\bq) +
V(q)\rho^{ind}(\bq)(1-G(q))]
\end{align}

\noindent where $G(q)$ is the local field factor. This results in:

\begin{align}
\label{eqn28b} \chi(q) = \frac{\chi^0(q)}{1-V(q)\chi^0(q)[1-G(q)]}
\end{align}

The local field factor can be
calculated in numerous ways.~\cite{Giuliani05}  In this work, we use
the Hubbard approximation \cite{HubbardPRSLA58,JonsonJPC76}, which
gives a local field factor:

\begin{equation}
\label{eqn29} G(q) = \frac{q}{2 \sqrt{q^{2} + k_{F}^{2}}}
\end{equation}

\noindent where $k_{F} = \sqrt{2 \pi p}$ is the Fermi wave-vector.
Although better approximations are available \cite{Giuliani05}, the
Hubbard approximation is sufficient to introduce a
density-dependence into the screening, unlike the Thomas-Fermi
approximation, which is density-independent.

\section{Comparison with other work on bilayer screening}

In this Appendix, we discuss how the analytical expression we obtain
for $\epsilon^{-1}_\nmi{2D,s,xf}(q,d)$ compares with other works on
linear screening theory for bilayer 2D systems produced in double
quantum well heterostructures, in particular, that of Zheng and
MacDonald~\cite{ZhengPRB94}. Note that we have translated the
equations from Ref.~21 into the notation used in our
paper for this Appendix.

Zheng and MacDonald begin by defining a density-density response
function (polarizability) $\chi_{ij}(q,w)$ for their bilayer 2D
system by:

\begin{align}
\label{eqn30} \rho_{i}(q,\omega) =
\sum_{j}\chi_{ij}(q,\omega)\phi_{j}^{ext}(q,\omega)
\end{align}

\noindent where $\rho$ is the linear density response (i.e., induced
charge density), $\phi_{j}^{ext}$ is the external potential, and
$i,j = 1,2$ are the layer indices with 1 being the transport layer
and 2 being the screening layer.  Zheng and MacDonald then use the
random phase approximation~\cite{BohmPR53} (RPA) and Singwi, Tosi,
Land and Sj\"{o}lander (STLS) approximation~\cite{SingwiPR68} to
obtain an expression for the polarizability:

\begin{widetext}
\begin{align}
\label{eqn31} \chi^{-1}(q,\omega) = \left(\begin{array}{cc}
[\chi^{0}_{1}(q,\omega)]^{-1} - V(q)[1 - G_{11}(q)] & U(q)[G_{12}(q) - 1] \\
U(q)[G_{21}(q) - 1] & [\chi^{0}_{2}(q,\omega)]^{-1} - V(q)[1 -
G_{22}(q)] \end{array}\right)
\end{align}
\end{widetext}

\noindent where $G_{ij}(q)$ are the local field factors that account
for the effects of exchange and correlation.~\cite{Giuliani05}  
For comparison with our work, we will consider $\omega = 0$, and ignore
interlayer exchange and correlations by setting $G_{12}(q) = G_{21}(q) = 0$, $G_{11}(q) =
G_{1}(q)$ and $G_{22}(q) = G_{2}(q)$. The latter approximation will be valid for large $d$,
but we would expect that $G_{ij}$ would become more significant at lower distances.
This is seen in the work of Liu \emph{et al.} ~\cite{LiuPRB96}, in which $G_{ii}$ and
$G_{ij}$ are calculated using STLS for different $d$, at $r_s = 4$.

In our work, we are seeking to obtain an effective single layer
dielectric function for the transport layer only.  Hence we only put
external charge density $\rho_{1}^{ext}(q)$ in the transport layer
and set the external charge density in the screening layer
$\rho_2^{ext}(q)$ to zero.  This results in external potentials in
the two layers of $\phi_{1}^{ext}(q) = V(q)\rho_{1}^{ext}(q)$, and
$\phi_{2}^{ext}(q) = U(q)\rho_{1}^{ext}(q)$. The total potential in
the transport layer can thus be expressed as:

\begin{align}
\label{eqn32} \phi_{1}(q) = \phi_{1}^{ext}(q) +
V(q)\rho_{1}^{ind}(q) + U(q)\rho_{2}^{ind}(q)
\end{align}

\noindent For the dielectric function of the transport layer, as
defined in Eqn.~\ref{eqn4}, this results in:

\begin{widetext}
\begin{align}
\label{eqn33} \frac{1}{\epsilon(q,d)} = 1 + V(q)\chi_{11}(q) +
U(q)\chi_{12}(q) + U(q)\chi_{21}(q) + e^{-qd}U(q)\chi_{22}(q)
\end{align}
\end{widetext}

\noindent This is analogous to 
Eqn.~\ref{eqn27} for the single layer case. Indeed, by applying $d
\rightarrow \infty$ to Eqn.~\ref{eqn33} reduces to Eqn.~\ref{eqn27}.
Finally, obtaining the matrix elements $\chi_{ij}(q)$ by inverting Eqn.~\ref{eqn31}
and inserting them into
Eqn.~\ref{eqn33}, we obtain the same expression as that given for
$\epsilon^{-1}_\nmi{2D,s,xf}(q)$ in Eqn.~\ref{eqn21} after returning
to zero thickness (i.e., $F(q)=1$).

\section{Comparison with the screening of a perpendicular electric field by a 2D system}

In this Section we show that the calculations in this paper, which describe the screening of in-plane point charges (and also arbitrary in-plane charge distributions) by a 2D system and also an adjacent ground plane, are consistent with the equations by Eisenstein \emph{et al.}\cite{EisensteinPRB94} describing the penetration of a perpendicular electric field across a 2D system.

\begin{figure}
\includegraphics[width=8.5cm]{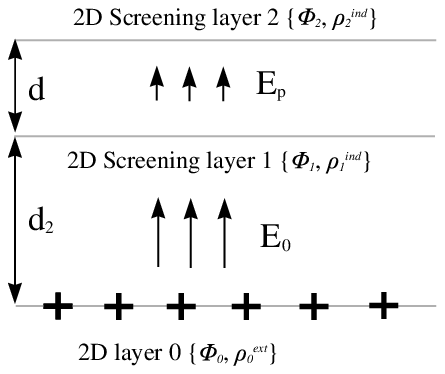}
\caption{\label{fig7}} Schematic showing the configuration used to calculate the penetration field across a 2D system. 
\end{figure}

In order to calculate the penetration of the perpendicular electric field, it is necessary to consider a slightly different configuration than previously used in Fig. \ref{fig2}(b). Figure \ref{fig7} shows a schematic of the system we now consider. We still have two 2D systems (labelled 1 and 2) separated by a distance \emph{d}. We now have no external charge in either of these layers, and only induced charges $\rho_{1}^{ind}$ and $\rho_{2}^{ind}$. In order to apply an electric field across layer 1, we place a layer of external charge $\rho_{0}^{ext}$  a distance $d_2$ below layer 1. The net electric fields $E_0$ and $E_p$ are shown in Fig. 7, and have been related by Eisenstein \emph{et al.} using Eqns. (5) and (6) in Ref. 34. Translating into the notation of our paper, and using $\chi^0_i= -e^2(\frac{\partial n}{\partial\mu})_i$, this results in:

\begin{align}
\label{eqn34} \frac{\partial E_p}{\partial E_0} = \frac{-\epsilon \chi^0_2}{- \epsilon (\chi^0_1 + \chi^0_2) + d \chi^0_1 \chi^0_2}
\end{align}

We now try to calculate the electric fields $E_0$ and $E_p$ directly using our model. The induced charges $\rho_1^{ind}$ and $\rho_2^{ind}$ are calculated in a similar fashion to previously in Section II. C., although Eqns. \ref{eqn18} and \ref{eqn19} need to be modified slightly to take into account the position of the external charge. This results in:

\begin{align}
\label{eqn35} \rho^{ind}_{1}(\bq) = \chi_{1}(\bq)
[\phi_1^{ext}(\bq) + U(\bq)\rho^{ind}_{2}(\bq)]\\
\label{eqn36} \rho^{ind}_{2}(\bq) = \chi_{2}(\bq)
[\phi_2^{ext}(\bq) + U(\bq)\rho^{ind}_{1}(\bq)]
\end{align}

\noindent where 

\begin{align}
\phi_1^{ext}(\bq)=V(\bq)e^{-\bq d_2}\rho_0^{ext}(\bq)\\
\phi_2^{ext}(\bq)=V(\bq)e^{-\bq(d+d_2)}\rho_0^{ext}(\bq).
\end{align}

We set $\rho_0^{ext}(\bf{r})=\sigma_0$, where $\sigma_0$ is a constant 2D surface charge density. This simplifies our model into the 1D problem considered by Eisenstein \emph{et al.}\cite{EisensteinPRB94} In \emph{q}-space this gives $\rho_0^{ext}(\bq)=\sigma_0 (2\pi)^2 \delta(\bq)$, where $\delta(\bq)$ is the Dirac delta function. Solving equations \ref{eqn35} and \ref{eqn36} simultaneously we obtain $\rho^{ind}_{i}(\bq) = \sigma_i (2\pi)^2 \delta(\bq)$ with:

\begin{align}
\label{eqn37} \sigma_{1} = \frac{\sigma_0 \chi^0_1 (\epsilon - d \chi^0_2)}
{-\epsilon(\chi^0_1 + \chi^0_2) + d \chi^0_1 \chi^0_2} \\
\label{eqn38} \sigma_{2} = \frac{\sigma_0 \epsilon  \chi^0_2}
{-\epsilon(\chi^0_1 + \chi^0_2) + d \chi^0_1 \chi^0_2},
\end{align}

\noindent where we have used the property of delta functions that $f(\bq)\delta (\bq) = f(\bf{0})\delta(\bq)$ for an arbitrary function $f(\bq)$. We note that the above equations \ref{eqn37} and \ref{eqn38} are valid even for an arbitrary local field correction, as at $q=0$ the Lindhardt function is equal to the Thomas-Fermi function, and for any self-consistent local field correction $G(0)=0$ (see equation 3.25 in Ref. 37.) Similarly, the equations \ref{eqn37} and \ref{eqn38} are valid even when allowing for the finite thickness of the 2D systems, since for an arbitrary form factor $F(q)$ we have $F(0)=1$.

We can now calculate the electric fields $E_0$ and $E_p$. For a single layer of uniform 2D charge density $\sigma$, the perpendicular electric field is given by $E= \frac{\sigma}{2 \epsilon}$. By considering the positions of the electric field $E_0$ and $E_p$ with respect to the charge layers $\rho_0$, $\rho_1$, and $\rho_2$, we obtain:

\begin{align}
\label{eqn39} E_p = \frac{\sigma_0 + \sigma_1 - \sigma_2}{2\epsilon}\\
\label{eqn40} E_0 = \frac{\sigma_0 - \sigma_1 - \sigma_2}{2\epsilon},
\end{align}

\noindent which results in

\begin{align}
\label{eqn41} \frac{E_p}{E_0} = \frac{-\epsilon \chi^0_2}{- \epsilon (\chi^0_1 + \chi^0_2) + d \chi^0_1 \chi^0_2}.
\end{align}

We note that in this model, the ratio $\frac{E_p}{E_0}$ is equal to $\frac{\partial E_p}{\partial E_0}$, as we have neglected other charges commonly present in 2D systems, such as regions of modulation doping. Our resulting equation \ref{eqn41} is thus in agreement with the equations used by Eisenstein \emph{et al.}\cite{EisensteinPRB94} (equation \ref{eqn34}).


\begin{thebibliography}:

\bibitem{WignerPR34} E. Wigner, Phys. Rev. {\bf46}, 1002 (1934).

\bibitem{CrandallPLA71} R.S. Crandall and R. Williams, Phys. Lett. A {\bf34}, 404 (1971).

\bibitem{GrimesPRL79} C.C. Grimes and G. Adams, Phys. Rev. Lett. {\bf42}, 795 (1979).

\bibitem{TsuiPRL82} D.C. Tsui, H.L. Stormer and A.C. Gossard, Phys. Rev. Lett. {\bf 48}, 1559 (1982).

\bibitem{LaughlinPRL83} R.B. Laughlin, Phys. Rev. Lett. {\bf50}, 1395 (1983).

\bibitem{AltshulerPhysE01} B.L. Altshuler, D.L. Maslov and V.M. Pudalov, Physica E {\bf9}, 2, (2001).

\bibitem{AbrahamsRMP01} E. Abrahams, S. V. Kravchenko and M. P. Sarachik, Rev. Mod. Phys. {\bf 73}, 251 (2001).

\bibitem{KravchenkoRPP04} S.V. Kravchenko and M.P. Sarachik, Rep. Prog. Phys., {\bf67}, 1, (2004).

\bibitem{PeetersPRB84} F.M. Peeters, Phys. Rev. B {\bf30}, 159 (1984).

\bibitem{WidomPRB88} A. Widom and R. Tao, Phys. Rev. B {\bf38}, 10787 (1988).

\bibitem{JiangSS88} H.-W. Jiang, M.A. Stan and A.J. Dahm, Surf. Sci. {\bf196}, 1 (1988).

\bibitem{MisturaPRB97} G. Mistura, T. G\"{u}nzler, S. Neser, and P. Leiderer, Phys. Rev. B {\bf56}, 8360 (1997).

\bibitem{HuangCM06} J. Huang, D.S. Novikov, D.C. Tsui, L.N. Pfeiffer and K.W. West, Cond-mat/0610320.

\bibitem{HoPRB08}  L.H. Ho, W.R. Clarke, A.P. Micolich, R. Danneau, O. Klochan, M.Y. Simmons, A.R. Hamilton, M. Pepper, and D.A. Ritchie, Phys. Rev. B {\bf77}, 201402(R) (2008).

\bibitem{Jackson99} J.D. Jackson, \emph{Classical Electrodynamics 3rd Ed.} (John Wiley and sons, New York, 1999).

\bibitem{Gradshteyn93} I.S. Gradshteyn and I.M. Ryzhik, \emph{Table of Integrals, Series and Products 5th Ed.}, (Academic Press, New York, 1993).

\bibitem{BohmPR53} D. Bohm and D. Pines, Phys. Rev. {\bf 92}, 609 (1953).

\bibitem{ThomasFermi27} L.H. Thomas, Proc. Camb. Philos. Soc. {\bf23}, 542 (1927); E. Fermi, Z. Phys. {\bf48}, 73 (1928).


\bibitem{SternPRL67} F. Stern, Phys. Rev. Lett. {\bf18}, 546 (1967).

\bibitem{Davies98} J.H. Davies, \emph{The Physics of Low-Dimensional Systems}, (Cambridge University Press, Cambridge 1998).

\bibitem{ZhengPRB94} L. Zheng and A.H. MacDonald, Phys. Rev. B {\bf49}, 5522 (1994).

\bibitem{SwierkowskiPRL91} L. \'{S}wierkowski, D. Neilson and J. Szyma\'{n}ski, Phys. Rev. Lett. {\bf67}, 240 (1991).

\bibitem{Giuliani05} G.F. Giuliani and G. Vignale, \emph{Quantum Theory of the Electron Liquid}, (Cambridge University Press, Cambridge 2005).

\bibitem{ShiPRL02} J. Shi and X.C. Xie, Phys. Rev. Lett. {\bf88}, 086401 (2002).

\bibitem{FoglerPRB2004} M.M. Fogler, Phys. Rev. B {\bf69}, 121409(R) (2004).

\bibitem{ZarembaPRL03} E. Zaremba, I. Nagy, and P. M. Echenique, Phys. Rev. Lett. {\bf90}, 046801 (2003).

\bibitem{HubbardPRSLA58} J. Hubbard, Proc. Roy. Soc. Lond. A {\bf 243}, 336 (1958).

\bibitem{JonsonJPC76} M. Jonson, J. Phys. C: Solid State Phys. {\bf9}, 3055 (1976).

\bibitem{AndoRMP82} T. Ando, A.B. Fowler and F. Stern, Rev. Mod. Phys. {\bf54}, 437 (1982).

\bibitem{SternJJAP74} F. Stern, Jap. J. Appl. Phys. Suppl. {\bf2(2)}, 323 (1974).

\bibitem{GoldPRB87} A. Gold, Phys. Rev. B {\bf35}, 723 (1987).

\bibitem{PricePRB84} P.J. Price, Phys. Rev. B {\bf30}, 2234 (1984).

\bibitem{TanatarPRB89} B. Tanatar and D. M. Ceperley, Phys. Rev. B {\bf39}, 5005 (1989).

\bibitem{EisensteinPRB94} J.P. Eisenstein, L.N. Pfeiffer and K.W. West, Phys. Rev. B {\bf50}, 1760 (1994).

\bibitem{SchakelPRB01} A.M.J. Schakel, Phys. Rev. B {\bf64}, 245101 (2001).

\bibitem{DolgovRMP81} O. V. Dolgov, D. A. Kirzhnits, and E. G. Maksimov, Rev. Mod. Phys. {\bf53}, 81 (1981)

\bibitem{IchimaruRMP82} S. Ichimaru, Rev. Mod. Phys. {\bf54}, 1017 (1982)

\bibitem{IwamotoPRB91} N. Iwamoto, Phys. Rev. B {\bf43}, 2174 (1991).

\bibitem{AdachiJAP82} S. Adachi, J. Appl. Phys. {\bf53}, 8775 (1982)

\bibitem{GoldammerJCG99} K. J. Goldammer, S. J. Chung, W. K. Liu, M. B. Santos, J. L. Hicks, S. Raymond and S. Q. Murphy, J. Crystal Growth, {\bf201-202}, 753(1999).

\bibitem{AllisonPRL06}  G. Allison, E. A. Galaktionov, A. K. Savchenko, S. S. Safonov, M. M. Fogler, M. Y. Simmons, and D. A. Ritchie, Phys. Rev. Lett. {\bf96}, 216407 (2006).

\bibitem{AshcroftSSP76} N.W. Ashcroft and N.D. Mermin, \emph{Solid State Physics} (Saunders College Publishing, Orlando, 1976).

\bibitem{SingwiPR68} K.S. Singwi, M.P. Tosi, R.H. Land and A. Sj\"{o}lander, Phys. Rev. {\bf 176}, 589 (1968).

\bibitem{Polini_Unpub} M. Polini, R. Asgari {\it et al.}, Manuscript in Preparation.

\bibitem{TracyCM08} L.A. Tracy, E.H. Hwang, K. Eng, G.A. Ten Eyck, E.P. Nordberg, K. Childs, M.S. Carroll, M.P. Lilly and S. Das Sarma, arXiv:0811.1394.

\bibitem{LiuPRB96} L. Liu, L. Swierkowski, D. Neilson, and J. Szyma\'{n}ski, Phys. Rev. B 53, 7923 (1996).

%
%





\end{thebibliography}
\end{document}